\documentclass[sn-basic,Numbered]{sn-jnl}

\usepackage{graphicx}%
\usepackage{multirow}%
\usepackage{amsmath,amssymb,amsfonts}%
\usepackage{amsthm}%
\usepackage{mathrsfs}%
\usepackage[title]{appendix}%
\usepackage{xcolor}%
\usepackage{textcomp}%
\usepackage{manyfoot}%
\usepackage{booktabs}%
\usepackage{algorithm}%
\usepackage{algorithmicx}%
\usepackage{algpseudocode}%
\usepackage{listings}%
\usepackage{eqnarray,amsmath,lipsum}
\usepackage{graphicx}
\usepackage{dcolumn}
\usepackage{mathtools}
\usepackage{soul,xcolor}
\usepackage{float}
\usepackage{color, soul}
\usepackage{bbold}
\usepackage[normalem]{ulem}
\setstcolor{red}
\usepackage[utf8]{inputenc}
\usepackage{natbib}

\theoremstyle{thmstyleone}%
\theoremstyle{thmstyletwo}%

\theoremstyle{thmstylethree}%

\begin{document}

\title[Universal and non-universal facets of quantum critical phenomena unveiled along the Schmidt decomposition theorem]{Universal and non-universal facets of quantum critical phenomena unveiled along the Schmidt decomposition theorem}


\author[1]{\fnm{Samuel M.} \sur{Soares}}
\equalcont{These authors contributed equally to this work.}

\author[1]{\fnm{Lucas} \sur{Squillante}}
\equalcont{These authors contributed equally to this work.}

\author[2]{\fnm{Henrique S.} \sur{Lima}}

\author[2,3,4]{\fnm{Constantino} \sur{Tsallis}}

\author*[1]{\fnm{Mariano} \sur{de Souza}}\email{mariano.souza@unesp.br}

\affil*[1]{\orgdiv{IGCE -- Physics Department}, \orgname{S\~ao Paulo State University (Unesp)}, \orgaddress{\city{Rio Claro}, \state{SP}, \country{Brazil}}}

\affil[2]{\orgname{Centro Brasileiro de Pesquisas F\'isicas}, \orgaddress{\street{Rua Xavier Sigaud 150}, \city{Rio de Janeiro}, \state{RJ}, \postcode{22290-180}, \country{Brazil}}}

\affil[3]{\orgname{Santa Fe Institute}, \orgaddress{\street{1399 Hyde Park Road}, \city{Santa Fe}, \state{NM}, \postcode{87501}, \country{USA}}}

\affil[4]{\orgname{Complexity Science Hub Vienna}, \orgaddress{\street{Metternichgasse 8}, \city{Vienna}, \postcode{1030}, \country{Austria}}}


\abstract{Critical phenomena have been extensively investigated both theoretically and experimentally in many fields, such as condensed matter physics, biology, e.g., brain criticality, and cosmology. In particular, the behaviour of response functions right at critical points (CPs) is highly topical. It turns out that in the frame of Boltzmann-Gibbs-von Neumann-Shannon approach, the extensive character of entropy breaks down at CPs. The latter implies diverging susceptibilities, which is at odds with experimental observations. Here, we investigate the influence of the spin magnitude $S$ on the quantum Gr\"uneisen parameter $\Gamma^{\textmd{0K}}_{q}$ right at CPs for the 1D Ising model under a transverse magnetic field. Our findings are fourfold: \emph{i)} for higher $S$, $\Gamma^{\textmd{0K}}_q$ is increased, but remains finite, reflecting the enhancement of the Hilbert space dimensionality; \emph{ii)} the Schmidt decomposition theorem recovers the extensivity of the nonadditive $q$-entropy $S_q$ only for a \emph{special} value of the entropic index $q$; \emph{iii)} the universality class in the frame of $S_q$ depends only on the symmetry of the system; \emph{iv)} we propose an experimental setup to explore finite-size effects in connection with the Hilbert space occupation at CPs. Our findings unveil both universal and non-universal aspects of quantum criticality in terms of $\Gamma^{\textmd{0K}}_{q}$ and $S_q$.}




\maketitle

\section{Introduction}\label{sec1}

The enhancement of thermodynamical response functions close to critical points (CPs) has been under a long debate \cite{Stanleybook}. Right at CPs, the extensivity of the Boltzmann-Gibbs entropy $S_{BG}$ breaks down \cite{prb2025, tsallisbook} and the system becomes nonergodic \cite{ergodic}. This is because the correlation length $\xi \rightarrow \infty$, giving rise to long-range correlations, which violate the probabilistic independence associated with $S_{BG}$ \cite{2q}. In other words, as stated by H.E. Stanley in Ref.\,\cite{Stanleybook}: \emph{``... closed form approximations have so far not been capable of describing the subtle and fascinating physical feature of the critical point---namely that although the interaction between the constituent magnetic moments is generally of extremely short range (for example, extending only to neighboring moments), this interaction nevertheless `propagates' from one moment to the next, tending to create a preferred direction for all magnetic moments and in fact the resulting order becomes infinite in range as the critical point is approached. ...''}. The same holds true regarding quantum critical phenomena, being the von Neumann entropy $S_N$ the counterpart of $S_{BG}$ \cite{vonNeumann, vonNeumannarticle}. The enhancement of thermodynamical response functions has been explored in various distinct situations, e.g., the CP in the supercooled phase of water \cite{supercooled}, and classical/quantum phase transitions \cite{griffiths, quantumGruneisen, prb2025}. It turns out that precisely at CPs, theoretical approaches in terms of $S_{BG}$ predict diverging thermodynamical response functions \cite{Stanleybook}. However, diverging physical quantities constitute an incongruent physical situation, cf.\,stated, for instance, by U.\,Harbach and S.\,Hossenfelder: \emph{``... The presence of infinities in physics always signals that we have missed some crucial point in our mathematical treatment. ...''} \cite{sabine}. In fact, the concept of ``infinity'' itself is widely debated in the literature as being a metaphorical concept and no conceivable test is able to support or reject it \cite{hess, infinity2025}. 
\begin{figure}[!t]
\centerline{\includegraphics[clip,width=1\columnwidth]{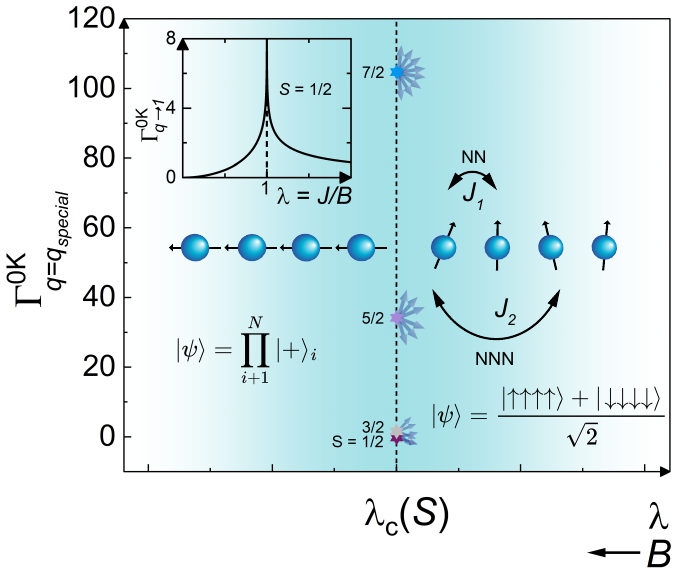}}
\caption{\footnotesize \textbf{Nondivergent quantum Gr\"uneisen parameter using $S_q$. a}, The quantum version of the Gr\"uneisen parameter in terms of $q$, $\Gamma^{\textmd{0K}}_{q=q_{special}}$, versus tuning parameter $\lambda = J/B$ for the \emph{special} value of the entropic index $q$, namely $q_{special}$, which ensures extensive $S_q$ right at the CP for the 1DIMTF. The vertical black dashed line indicates the critical value of $\lambda$, namely $\lambda_c$, which is $S$-dependent (Supplementary Information, section\,1). The interaction between NN spins is represented by $J_1$ and NNN by $J_2$. For $B \rightarrow  0$, i.e., $\lambda \rightarrow \infty$, the ground state wave function is a superposition of $|\!\!\uparrow\uparrow\uparrow\uparrow\rangle$ and $|\!\!\downarrow\downarrow\downarrow\downarrow\rangle$ states. For $B \rightarrow \infty$ $(\lambda \rightarrow 0)$, the system is in the so-called quantum paramagnetic phase, i.e., all spins are aligned along the direction of $B$ and the system is described as a product state. For simplicity, we have considered only four spins in this scheme. The stars in different colours indicate the value of $\Gamma^{\textmd{0K}}_{q=q_{special}}$ for each value of $S$ right at $\lambda_c$ considering the system size $N = 8$. The possible spin projections are also indicated for each value of $S$ by arrows in different colours. \textbf{b}, $\Gamma^{\textmd{0K}}_{q \rightarrow 1}$ versus $\lambda$, considering NN interaction for $S = 1/2$ in the thermodynamic limit $(N \rightarrow \infty)$ for comparison \cite{quantumGruneisen}. \textbf{c}, The nonadditive $q$-entropy $S_q$ (left axis) for various values of $q$ and $\Gamma^{\textmd{0K}}_{q\rightarrow 1}$ (right axis) versus $\lambda$ considering $N = 11$ for $S = 3/2$. The yellow bullet represents the inflection point, while the black solid lines in each curve indicate $dS_q/d\lambda$ in the vicinity of the CP. The vertical black dashed line indicates the critical value of $\lambda$, i.e., $\lambda_c = 0.46$ (Supplementary Information, section\,1).}
\label{Fig-1}
\end{figure}
In this context, an appropriate parameter to explore critical phenomena at finite temperature $T$ is the so-called Gr\"uneisen ratio $\Gamma$, being its quantum version $\Gamma^{0\textmd{K}}_{q\rightarrow 1}$ proposed by some of us \cite{quantumGruneisen}, where $q$ is the so-called entropic index. In Ref.\,\cite{quantumGruneisen}, using the 1D Ising model under a transverse magnetic field (1DIMTF) as a case of study, we have shown that right at a genuine quantum CP, $\Gamma^{0\textmd{K}}_{q \rightarrow 1}$ is enhanced in the frame of $S_N$, cf.\,Fig.\,\ref{Fig-1}\,b. It turns out that right at the CP, $S_N$ follows a logarithmic dependence with the system size $N$ \cite{osterloh, kitaev, korepin}, leading to a divergent
behaviour of $\Gamma^{0\textmd{K}}_{q \rightarrow 1}$ in the thermodynamic limit \cite{quantumGruneisen}. Recently, we have proposed a way of regularizing the critical phenomena theory right at CPs on the ground of the nonadditive $q$-entropy $S_q$ \cite{prb2025,tsallistop} by 
illustrating with the 1DIMTF. We have shown that for a \emph{special} value of the entropic index $q$, namely $q_{special}$, which is system dependent and ensures extensive $S_q$, $\Gamma^{\textmd{0K}}_{q=q_{special}}$ is universally nondivergent right at CPs even in the thermodynamic limit \cite{prb2025}. Here, we demonstrate that the extensivity of $S_q$ emerges naturally for $q = q_{special}$ in the frame of the Schmidt decomposition theorem (SDT), highlighting universal and non-universal aspects of quantum criticality. Yet, we propose an experimental setup for exploring finite-size effects. We explore the 1DIMTF CP in the frame of $S_q$ considering interaction between nearest neighbors (NN) for $S = 3/2$, 5/2, 7/2 and next NN (NNN) for $S = 1/2$, aiming to unveil the influence of the $S$ magnitude, i.e., the number of spin projections, on $\Gamma^{\textmd{0K}}_{q=q_{special}}$ right at the CP, cf.\,Fig.\,\ref{Fig-1}\,a. 

\section{Results, analysis, and discussions} Before starting our discussions, it is worth recalling that considering the 1DIMTF for $S = 1/2$, the CP takes place at $\lambda = J/B = 1$ \cite{pfeuty}, where $J$ is the magnetic coupling constant between NN spins and $B$ the modulus of the transverse magnetic field. However, the critical value of $\lambda$, i.e., $\lambda_c$, depends on the considered spin operator $\hat{S}$ \cite{multifractal, spindependence}. In other words, the critical magnetic field $B_c$ needed to induce a transition from a ferromagnetic to a quantum paramagnetic phase depends on $S$, which in turn is one of the quantum numbers associated with $\hat{S}$ \cite{zettili}. For vanishing $B$ $(\lambda \gg \lambda_c)$, the system is in a ferromagnetic phase, i.e., it lies in a superposition of spin up and down states \cite{sachdev, sachdevIsing}, cf.\,Fig.\,\ref{Fig-1}\,a. For $B \rightarrow \infty$ $(\lambda \ll \lambda_c)$, all spins are aligned parallel to the applied field, the so-called quantum paramagnetic phase. Right at $\lambda = \lambda_c$, all spins are entangled and hence the state is given by $|\psi\rangle = \sum_{p=1}^{P}\sum_{k=1}^{K}c_{p,k}|\chi_A^p\rangle\otimes|\xi_B^k\rangle$, where $c_{p,k}$ are the coefficients associated with each tensor product, $|\chi_A^p\rangle$ and $|\xi_B^k\rangle$ the orthonormalized kets, $p$ and $k$ the summation indices over the orthonormal bases, $P$ and $K$ the dimension of each ket, and $A$ and $B$ the labels of each subsystem \cite{sachdev, sachdevIsing, cohen3}. Due to the high degeneracy and since the spins are highly entangled right at the CP, the number of accessible states increases. Such a feature is captured by $S_q(\hat{\rho})$, which is more sensitive to subtle variations of $\lambda$ in the critical regime. In what follows, we explore the 1DIMTF in the frame of $S_q$ considering NN interaction for distinct values of $S\,> 1/2$, which implies in additional spin projections. For $S = 1/2$, we go beyond our previous work \cite{prb2025} by considering NNN interaction. Upon considering the generalized ratio $\lambda = g/h$, where $g$ and $h$ are two distinct tuning parameters, e.g., $B$, or a given energy scale, it suffices to analyze solely $\Gamma^{\textmd{0K}}_{q} = dS_q/d\lambda$ \cite{prb2025}, cf.\,adopted in this work.

\subsection{The 1DIMTF for distinct $S$ values} 
\begin{figure}[!t]
\centerline{\includegraphics[clip,width=1\columnwidth]{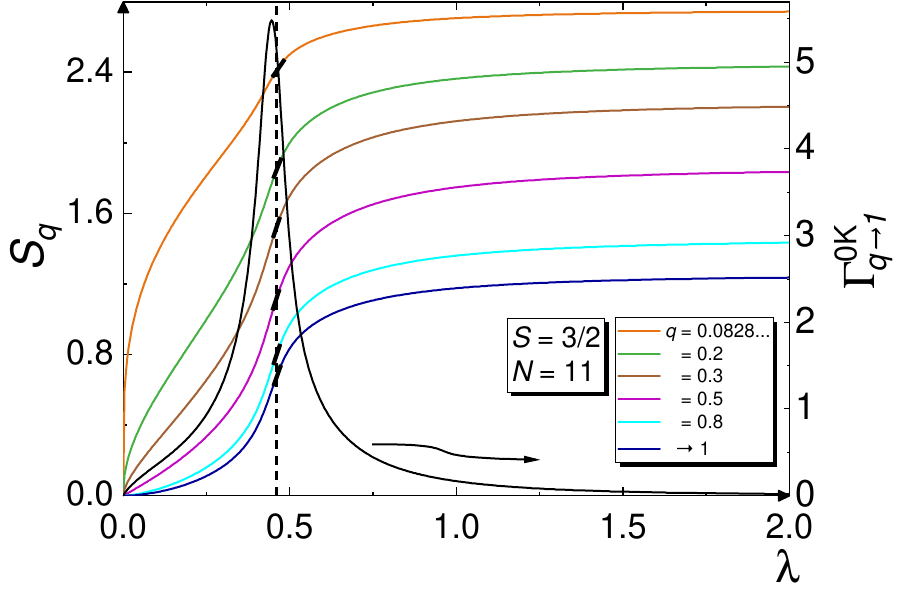}}
\caption{\footnotesize \textbf{Nondivergent quantum Gr\"uneisen parameter in the thermodynamic limit at CPs. a}, $(\Gamma^{\textmd{0K}}_{q=q_{special}})^{-1}$ versus $[1/\ln_q(N)]^{\textmd{a}_q}$ for $q = 0.0828\ldots$ and $\lambda = \lambda_c$ considering spin $S = 1/2$, 3/2, 5/2, and 7/2. As the value of $S$ increases, the extrapolation of $\Gamma^{\textmd{0K}}_{q=q_{special}}$ for $N\rightarrow\infty$ is also enhanced, cf.\,indicated by stars in different colours. \textbf{b}, $(\Gamma^{\textmd{0K}}_{q=q_{special}})^{-1}$ versus $[1/\ln_q(N)]^{\textmd{a}_q}$ for $q = 0.0828\ldots$ and $S = 1/2$ considering NN and NNN interaction. For $S = 1/2$ with NN interaction $\textmd{a}_q = 0.5$, while considering NNN interaction $\textmd{a}_q = 1.32$. For the cases of NN and NNN interactions, $N$ ranges from 2 to 26 and from 2 to 14, respectively. \textbf{c}, 
$\lim_{N \rightarrow \infty}\Gamma^{\textmd{0K}}_{q=q_{special}}$ versus $S$ for $q = 0.0828\ldots$ and $\lambda = \lambda_c$. Given computational limitations, we were able to obtain data set up to $N = 8$. We verify that $\Gamma^{\textmd{0K}}_{q}(S) = C\left[S(S+1)\right]^{\beta}$, where $C=0.43$ and $\beta=2$ are fitting parameters. \textbf{d}, $\lim_{N\rightarrow \infty}\Gamma^{\textmd{0K}}_{q=q_{special}}$ versus spin $S$ for $S = 1/2$ considering NN and NNN interaction. The data set depicted in all panels were computed for $\lambda = \lambda_c$.} 
\label{Fig-2}
\end{figure}

We begin our analysis by recalling the $q$-generalized version of the Boltzmann-Gibbs-von Neumann-Shannon entropy $S_q(\hat{\rho})$, namely \cite{tsallisbook, TsallisCaruso}
\begin{equation}
S_q(\hat{\rho}) = k\frac{1 - \textmd{Tr}\hat{\rho}^q}{q - 1},
\label{SN}
\end{equation}
\noindent where $k$ is a positive constant, and $\hat{\rho}$ the density matrix operator, being that
$q \rightarrow 1$ $\Rightarrow$ $S_1 = S_N$ \cite{tsallisbook, TsallisCaruso}. Following discussions in Ref.\,\cite{prb2025}, $\Gamma^{\textmd{0K}}_{q\rightarrow 1}$ can be rewritten straightforwardly in terms of $S_q(\hat{\rho})$.
Essentially, to obtain $S_q(\hat{\rho})$ we need first to compute $\hat{\rho} = \sum_np_n|\psi_n\rangle\langle\psi_n|$, where $p_n$ are the probabilities of each state $|\psi_n \rangle$ obtained through the diagonalization of the Hamiltonian $\hat{H}$ \cite{quantumGruneisen}. To this end, we recall $\hat{H}$ of the 1DIMTF \cite{spin1, pfeuty, TsallisAndre}
\begin{equation}
\hat{H} = -J\sum_{\langle i,j \rangle}^{N-1}\hat{S}_{i}^z\hat{S}_{j}^z - B\sum_{i=1}^{N}\hat{S}_{i}^x,
\label{HS132}
\end{equation}
\noindent where $\hat{S}_i^z$ and $\hat{S}_j^z$ are the spin operators at, respectively, the $i$th- and $j$th-site along the $z$-axis, $\hat{S}_i^x$ the spin operator at the $i$th-site along the $x$-axis, and $N$ the total number of spins. \noindent According to Ref.\,\cite{TsallisCaruso}, for all systems belonging to the Ising universality class, $q=\sqrt{37} - 6 = 0.0828\ldots$ ensures the extensivity of $S_q$ right at the CP. It is evident that upon increasing $S$, a larger number of spin configurations is accessible in comparison with those for $S = 1/2$ reported in Ref.\,\cite{prb2025}. Thus, the $\lambda$ derivative of $S_q$, i.e., $\Gamma^{\textmd{0K}}_q$, at the CP is more sensitive to minute variations of $\lambda$, cf.\,Fig.\,\ref{Fig-1}\,b for $S = 3/2$. It is worth mentioning that we have computed the single-site reduced density matrix $\hat{\rho}$ for the data set depicted in Figs.\,\ref{Fig-1}, \ref{Fig-2}\,a and c.
Note that the maximum of $\Gamma^{\textmd{0K}}_{q\rightarrow 1}$ shifts towards $\lambda_c = 0.46$ as $N$ increases, cf.\,Fig.\,S1 (Supplementary Information, section\,1).  
Figure\,\ref{Fig-2}\,a depicts $(\Gamma^{\textmd{0K}}_{q=q_{special}})^{-1}$ versus $[1/\ln_q(N)]^{\textmd{a}_q}$, being $\ln_q(N) = (N^{1-q}-1)/(1-q)$ the $q$-logarithmic function \cite{tsallistop} and $\textmd{a}_q$ a $q$-dependent fitting parameter, for $S = 1/2$ \cite{prb2025}, 3/2, 5/2, and 7/2 considering NN interaction for $q=0.0828\ldots$. It is to be noted that upon extrapolating our results for $N \rightarrow \infty$, each data set intercepts the ordinate axis at different values, i.e., $\Gamma^{\textmd{0K}}_{q=q_{special}}$ for $N \rightarrow \infty$ is distinct for each $S$ considered, cf.\,Figs.\,\ref{Fig-1} and \ref{Fig-2}\,a. This remarkable feature is associated with how $S_q$ varies with respect to $\lambda$ right at the CP for each case, because, as previously discussed, increasing the value of $S$ also increases the number of accessible states. Furthermore, it is to be noted that the difference between $(\Gamma^{\textmd{0K}}_{q=q_{special}})^{-1}$ for $S = 1/2$, 3/2, 5/2, and 7/2 is related with the number of elements of $\hat{\rho}$, which is embedded in $S_q({\hat{\rho}})$, cf.\,Eq.\,\ref{SN}. Considering a system of $N$ spins, the total Hilbert space has dimension $(2S + 1)^N$ \cite{sakurai}, which in turn is reflected in $\hat{\rho}$. Thus, the expressive difference between the values of $(\Gamma^{\textmd{0K}}_{q=q_{special}})^{-1}$ for distinct spin projections is associated with how $S_q$ scales with $S$. In other words, as the value of $S$ is increased, more degrees of freedom emerge in the system \cite{matrixrepresentation}, which in turn are embodied in $\hat{\rho}$, captured by $S_q(\hat{\rho})$, and reflected in $\Gamma^{\textmd{0K}}_{q}$. Following Refs.\,\cite{cohen3, sakurai}, considering a system composed by two nonentangled spins $A$ and $B$ with $S = 1/2$, its Hilbert space is given by $\mathcal{H}_{AB} = \mathcal{H}_A \otimes \mathcal{H}_B$, where each space $\mathcal{H}_A$ and $\mathcal{H}_B$ has dimension $2$ \cite{lebellac, hilbertspace}. Following previous discussions, for $N$ spins the dimensionality of $\mathcal{H}$ is given by $(2S + 1)^N$ \cite{sakurai}. As a consequence, the density matrix $\hat{\rho}$ has $(2S + 1)^N \times (2S + 1)^N$ elements. As $S$ increases, the Hilbert space dimensionality grows, leading to an enhancement of $\Gamma^{\textmd{0K}}_{q}$. Upon considering that the two spins are entangled, the state that describes the total system is not a unique tensor product, but a linear combination of tensor products. Such entangled states lead to a higher number of accessible states, so that the number of Eigenvalues of $\hat{\rho}$ which will contribute for the computation of $S_q(\hat{\rho})$ will increase, leading to a higher value of $\Gamma^{\textmd{0K}}_{q}$. Therefore, $\Gamma^{\textmd{0K}}_{q}$ can be considered as a \emph{tool} to probe how the Hilbert space evolves as a function of $S$, $N$, and $\lambda$. Figure\,\ref{Fig-2}\,b shows $(\Gamma^{\textmd{0K}}_{q=q_{special}})^{-1}$ versus $[1/\ln_q(N)]^{\textmd{a}_q}$ for $S = 1/2$ and $q = 0.0828\ldots$ considering NN and NNN. Figure\,\ref{Fig-2}\,c depicts $\lim_{N \rightarrow \infty}\Gamma^{\textmd{0K}}_{q=q_{special}}$ versus $S$ for $q = 0.0828\ldots$ right at the CP, namely $\lambda = \lambda_c$. For increasing $S$, $\Gamma^{\textmd{0K}}_{q=q_{special}}$ is also enhanced. Our findings demonstrate the $S$-dependence of $\Gamma^{\textmd{0K}}_{q=q_{special}}$ right at the CP, cf.\,Figs.\,\ref{Fig-2}\,a and \ref{Fig-2}\,c, which is one of the key findings of the present work. We associate the nonlinear $S$-dependence of $\lim_{N \rightarrow \infty}\Gamma^{\textmd{0K}}_{q=q_{special}}$, cf.\,Fig.\,\ref{Fig-2}\,d, with the nonlinear scaling of $\hat{\rho}$ with $N$. 

\subsection{The 1DIMTF considering NNN interaction for $S = 1/2$ under the light of $S_q$} The Hamiltonian of the 1DIMTF considering NNN interaction is given by \cite{NNN}
\begin{equation}
\hat{H} = -J_1 \sum_{i=1}^{N}\hat{S}_i^z\hat{S}_{i+1}^z-J_2 \sum_{i=1}^{N}\hat{S}_i^z\hat{S}_{i+2}^z - B\sum_{i=1}^{N}\hat{S}_i^x,
\end{equation}
\noindent where $J_1$ and $J_2$ are, respectively, the coupling constants between NN and NNN, $\hat{S}_{i+1}^z$ and $\hat{S}_{i+2}^z$ the spin operators in the ($i$th+1)- and ($i$th+2)-sites, respectively, along the $z$-axis. Following discussions in Ref.\,\cite{NNN}, $B_c$ depends on the ratio $J_2/J_1$, being considered in the present work $-J_2/J_1 = 0.32$, which corresponds to $B_c/J_1 = 0.3939$ \cite{NNN}. It is to be noted that also for $S = 1/2$ with NNN, $q \rightarrow 1$ and $N \rightarrow \infty \implies \Gamma^{\textmd{0K}}_{q=q_{special}} \rightarrow \infty$ right at the CP, cf.\,Fig.\,S8 (Supplementary Information, section\,2). For $q = 0.0828\ldots$, $\Gamma^{\textmd{0K}}_{q=q_{special}}$ remains finite at the CP even in the thermodynamic limit, cf.\,Fig.\,\ref{Fig-2}\,d. We stress that we have investigated only NNN for $S = 1/2$ because the value of $B_c$ depends on the ratio $J_2/J_1$ \cite{NNN}. However, to the best of our knowledge, this ratio has not yet been reported in the literature for higher values of $S$, and therefore the exact region in the phase diagram where the CP lies remains to be determined. Our results regarding the $S$-dependence of $\Gamma^{\textmd{0K}}_{q=q_{special}}$ right at the CP, considering both NN and NNN demonstrate that systems with Ising-type symmetry belong to the same universality class, which constitutes an important result of the present work. For the \emph{XY} model, for instance, the universality class is fixed for $q_{special} = \sqrt{10} - 3$ \cite{TsallisCaruso}. Next, we discuss the SDT
in terms of $S_q$.

\subsection{The nonadditive entropy $S_q$ and the Schmidt decomposition theorem} 
\begin{figure}[!t]
\centerline{\includegraphics[clip,width=1\columnwidth]{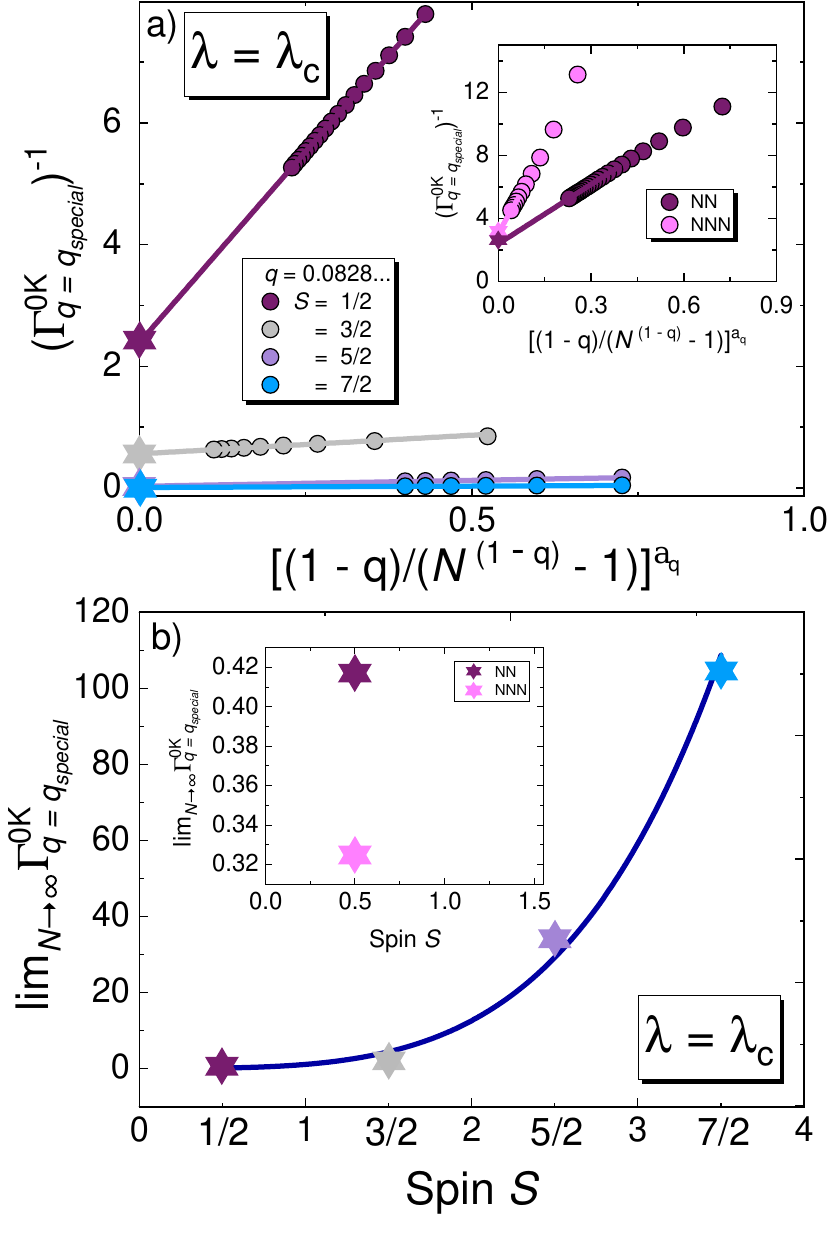}}
\caption{\footnotesize \textbf{The extensivity of $S_q$ and the Schmidt decomposition theorem. a}, Schematic representation of the 1DIMTF right at the CP with $N$ spins and block size $L$. We consider $L = 2$ for simplicity. The states $|u_i\rangle$ and $|w_i\rangle$ describe the subsystem with $L$ and $(N - L)$ spins, respectively. The magnifying glass indicates that the $q$-entropy was employed to explore the system. \textbf{b}, The $q$-generalized entropy $S_q$ versus block size $L$ for $q = 0.01$, \textbf{c} $q = q_{special} = 0.0828\ldots$, and \textbf{d} $q = 0.7$ for various values of $N$, considering $S = 1/2$ and $\lambda = 1$. The black solid line in panel \textbf{c} represents the linear fit employed to analyze $S_q$ in the regime of $L \leq N/2$. \textbf{e}, Scheme of how to attain high fidelity when transmitting information from A to B. The calculation sequence starting from $\hat{H}$ to $S_q$ is indicated. Details in the main text.}
\label{Fig-3}
\end{figure}

The so-called SDT enables one to write the state  $|\psi \rangle$ of a bipartite system composed by subsystems $A$ and $B$ as $|\psi \rangle = \sum_{i}^{}\lambda_i|u_i\rangle \otimes |w_i\rangle$, being $\lambda_i$ the probability that satisfies $\sum_{i}^{} \lambda_i^2 = 1$, and $|u_i\rangle$ and $|w_i\rangle$ the set of orthonormal states describing $A$ and $B$, respectively \cite{schmidtoriginal, quantumcomputation, cohen3}. The Eigenvalues of the density matrix operators $\hat{\rho}^A$ and $\hat{\rho}^B$ are the same, since entanglement is symmetric in terms of $A$ and $B$ \cite{cohen3}. As a consequence, $S_N$ of both systems is also the same, since it is fully determined by the Eigenvalues of $\hat{\rho}$ \cite{quantumcomputation, sakurai}. The SDT has been applied to entanglement characterization \cite{entanglementschmidt}, thermodynamics of arbitrary bipartite systems \cite{thermoschmidt}, and for the entropy of Schwarzschild-type black holes \cite{schmidtblackhole}, just to mention a few examples. Here, we analyze the spectrum of $\hat{\rho}$ aiming to explore finite-size effects upon employing exact diagonalization. In order to analyze the SDT in terms of $S_q$, we consider a spin chain with $N$ spins and then, for a fixed block $L$ and the rest of the chain $(N-L)$, we compute the entanglement, i.e., $S_q$, between $L$ and $(N-L)$, cf.\,Fig.\,\ref{Fig-3}\,a. 
Then, we consider $(L+1)$ in the computation of $S_q$ and so on up to $L = (N-1)$, so that we are able to obtain $S_q$ versus $L$ for distinct values of $N$ and $q$. Regarding Figs.\,\ref{Fig-3}\,b-d, $S_q$ versus $L$ is depicted for $S = 1/2$ and various values of $N$ considering $\lambda = 1$ and $q = 0.01$, 0.0828..., and 0.7. For $L < N/2$, $S_q$ is enhanced upon increasing $L$ and maximized for $L = N/2$. For $N \rightarrow \infty$, the maximum in $S_q$ versus $L$ is shifted towards $L \rightarrow \infty$ because $S_q$ is symmetric at $L = N/2$. It is remarkable that only for $q = q_{special}$, the extensivity of $S_q$ right at the CP is recovered \cite{TsallisCaruso}. This is a key result of the present work and can be naturally extended to distinct values of $S$, being the value of $q_{special}$ the same for any $S$, cf.\,Figs.\,S9, S10, S11, S13 for NN and Fig.\,S12 for NNN (Supplementary Information, section\,3). This is because the universality class is the same, since the Ising-type symmetry is maintained. Hence, our findings demonstrate that only for $q = 0.0828\ldots$, $S_q \propto L$, corroborating the results reported in Refs.\,\cite{TsallisCaruso, prb2025}, i.e., $q = 0.0828\ldots$ is \emph{special} because it recovers the extensivity of $S_q$ right at the CP, 
as mandated by the Legendre structure of Thermodynamics
\cite{TsallisCaruso,prb2025}. This is remarkable because employing $S_q$ 
for $q = q_{special}$, the diverging susceptibilities right at CPs, a fingerprint of the Boltzmann-Gibbs statistical mechanics, are regularized \cite{prb2025, chaostsallis}, cf.\,Figs.\,\ref{Fig-2}\,a and c. The present analysis regarding the linear behaviour of $S_q$ versus $L$ in terms of the SDT is not to be confused with the one reported in Ref.\,\cite{vidalprl}. There, the authors consider an expansion of $|\psi\rangle$ and its coefficients, obtaining a linear relation between the number of qubits $n$ and entanglement $E_{\chi}$. Next, we discuss the relevance of such results for quantum computing. A key requirement for quantum computation is that the entanglement scales linearly with $N$ \cite{vidalprl}. This is because the entropy must reflect how many quantum states are effectively available to the system, since this determines how much information can be faithfully represented \cite{quantumcoding}. If this number is under or over-estimated when computing the entropy, the recovered information is no longer reliable, i.e., fidelity is enhanced for $q = q_{special}$, cf.\,Fig.\,\ref{Fig-3}\,e. Considering quantum annealing-type architectures, the system necessarily passes through the CP, where extensivity of the entropy breaks down using Boltzmann-Gibbs-von Neumann-Shannon entropy. Our findings demonstrate that the diverging susceptibilities ``pathologies'' at CPs are not intrinsic. Instead, their origin come from making use an unappropriate entropy approach, cf.\,Fig.\,\ref{Fig-3}\,e. Hence, our findings may serve as a basis for optimizing both the computational cost and time to simulate quantum many-body systems \cite{vidalprl}. Yet, our findings might be relevant for quantum machines \cite{quantumrefrigerator}, since the mathematical structure of Thermodynamics is 
preserved upon considering $S_{q = q_{special}}$.\newline
\subsection{Probing the Hilbert space at CPs}
Classical textbooks discuss the Hilbert space, which typically contains infinite accessible states \cite{lebellac, sakurai, hilbertspace, cohen3, zettili, griffithsbook}. The infinite dimensionality of the Hilbert space is associated with the Heisenberg's canonical commutation relation. Upon taking the trace of the commutator, which is always equal to zero \cite{HSinfinity}, between the momentum operator $\hat{P}$ and the position operator $\hat{Q}$, one finds that $\textmd{Tr}[\hat{P},\hat{Q}] = -i\hbar\textmd{Tr}\hat{\mathbb{1}} \neq 0$, where $\hat{\mathbb{1}}$ is the identity matrix operator. Nevertheless, the trace is an undefined operation in an infinite space, so that one must define an unbounded Hilbert space to maintain the physical consistency of the commutation relations \cite{lebellac, HSinfinity}. However, in reality, the number of accessible states is limited to the particular system of interest. It turns out that even at CPs and in the thermodynamic limit, the ways in which a system can be arranged are finite. Based on our previous analysis, upon increasing $S$, the number of accessible states is enhanced, which is reflected in $\hat{\rho}$. Essentially, the occupation of the Hilbert space is strongly $\lambda$-dependent, so that for $\lambda_c$ the number of available states in the Hilbert space is the highest \cite{prb2025}. Considering $q > q_{special}$, the number of accessible states are over-estimated, which leads to a divergent-like behaviour of $\Gamma^{\textmd{0K}}_{q}$ \cite{quantumGruneisen}. For $q < q_{special}$, the states that are actually accessible to the system are under-estimated for the computation of $S_q(\hat{\rho})$, leading to $\Gamma^{\textmd{0K}}_{q} \rightarrow 0$. It turns out that only for $q = q_{special}$ the number of accessible states is properly taken into account in the frame of $S_q$. In other words, our approach enables us to take into account properly how the system ``visits'' the Hilbert space. Note that $\Gamma^{\textmd{0K}}_{q}$ captures the enhancement of the Hilbert space population at the CP upon increasing the number of accessible states of the system. Thus, we recognize $\Gamma^{\textmd{0K}}_{q}$ as a proper \emph{tool} to access the actual Hilbert space occupation. Our analysis is not to be confused with the concept of Hilbert space fragmentation \cite{fragmentation}.\newline
\section{A proposal to probe experimentally finite-size effects right at finite temperature CPs} In order to explore finite-size effects in real systems, it is necessary to vary the size of the system in the region of interest in the phase diagram. To this end, we propose a photostrictive experiment in which the incident light spot size $L$ can be systematically varied, so that we are able to simulate the variation of the size of the system, i.e., the block size, cf.\,Fig.\,\ref{Fig-3}\,a. 
\begin{figure}[!t]
\centerline{\includegraphics[clip,width=0.55\columnwidth]{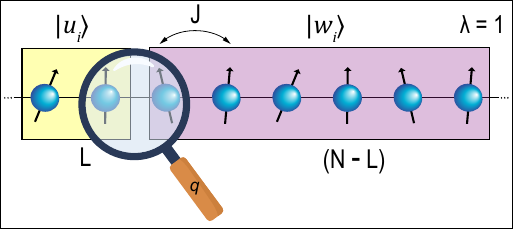}}
\caption{\footnotesize \textbf{Finite-size effects in real systems.} Schematic representation of our proposal to explore finite-size effects right at CPs showing the investigated specimen and the monochromatic light source. Upon increasing the spot size $L$ of the incident light, more electric dipoles will be enclosed and thus contribute for the photostriction effect, enabling the exploration of size effects.}
\label{Fig-4}
\end{figure}
In other words, we propose to systematically measure the photostriction coefficient upon increasing the spot size. Hence, by considering a globally entangled system, only a portion of it would contribute to the photostrictive effect given a particular spot size. Also, to the best of our knowledge, a magnetic system which simulates the 1DIMTF and presents photostrictive effects is not reported in the literature up to date. Thus, in order to overcome the difficulties posed in exploring a magnetic system with such characteristics, we propose the investigation of a ferroelectric system which possess the key ingredients to explore our proposal. In this context, the system SbSI constitutes a good candidate for this proposal, since it presents a ferroelectric transition with a Curie point at 295\,K \cite{SbSI, SbSI1962}. In our proposal, upon systematically varying $L$, we can simulate a variation of $N$ since a larger number of electric dipoles will be enclosed in the light spot upon increasing $L$, cf.\,Fig.\,\ref{Fig-4}. It is clear that we are dealing with a significant size of the system even for a laser whose spot diameter is typically of a few micrometers. However, increasing the spot size to a millimeter scale can give insights to finite-size effects accessible in the proposed experiment. For instance, upon increasing the spot size from $\mu$m to mm and considering that the typical distance between charges in an electric dipole is on the order of 0.1\,\AA \,\cite{electricdipole}, the number of electric dipoles that will be enclosed changes from $\sim$10$^{10}$ to $\sim$$10^{16}$. It is clear that our theoretical analysis in previous subsections regarding the 1DIMTF for additional spin projections is genuine quantum, i.e., $T = 0$\,K, which cannot be achieved in the real world \cite{unveiling}. It would be highly desirable to have a system that presents photostrictive effects in the mK range, aiming to explore quantum critical manifestations of matter.

\section{Quintessence, conclusions, and outlook} We have investigated the 1DIMTF for $S = 3/2$, 5/2, 7/2, as well as considering NNN for $S=1/2$ at the CP using $S_q$. For $q = q_{special}$, $\Gamma^{\textmd{0K}}_{q=q_{special}}$ remains finite even in the thermodynamic limit, being distinct for each value of $S$ as direct consequence of the distinct number of spin projections. 
Also, the breakdown of the Hellmann-Feynman theorem is regularized at the CP upon considering $q = q_{special}$ \cite{prb2025, quantumGruneisen}. We have analyzed the SDT in terms of $S_q$ and demonstrated that the extensivity of $S_q$ is recovered solely upon considering $q = q_{special}$.  Our findings linking the extensivity of $S_q$ with SDT and the occupation of the Hilbert space at CPs open a route that might be relevant to the field of quantum computing \cite{vidalprl}. A formal mathematical $q$-generalization of the SDT would be very welcome. Our results regarding quantum computing are conceptual rather than algorithmic. We provide a framework based on $S_q$ and $\Gamma^{\textmd{0K}}_q$ that correctly quantifies entanglement and susceptibilities at CPs. Our approach might be relevant for adiabatic and annealing-based quantum computing architectures. We report on universal and non-universal aspects of quantum critical phenomena, which includes the $S_q$ and $\Gamma^{\textmd{0K}}_q$ dependencies on $N$, $S$, the range of interactions, and the universality class of the investigated model. Essentially, for systems presenting Ising-type symmetry, $q_{special}$ remains the same for any $N$ and $S$. Our analysis can be extended to other fields, such as dynamical systems and situations where long-range forces govern the system, as well as the exploration of the bosonic case, which constitute part of ongoing projects. Yet, we have proposed an experimental setup to explore finite-size effects in real systems employing photostriction measurements.

\section{Acknowledgements} MdeS acknowledges partial financial support from the S\~ao Paulo Research Foundation -- Fapesp (Grants 2011/22050-4, 2017/07845-7, and 2019/24696-0), National Council of Technological and Scientific Development -- CNPq (Grants 303772/2023-9), and Capes -- Finance Code 001 (M.\,Sc.\,fellowship of SMS). CT acknowledges partial financial support from National Council of Technological and Scientific Development--CNPq and Fundac\~ao Carlos Chagas Filho de Amparo \`a Pesquisa do Estado do Rio de Janeiro -- Faperj. We also acknowledge Laborat\'orio Nacional de Computa\c{c}\~ao Cient\'ifica-LNCC (Brazil) for allowing us to use the Santos Dumont (SDumont) supercomputer. This research was supported by resources supplied by the Center for Scientific Computing (NCC/GridUNESP) of the S\~ao Paulo State University (UNESP). SMS and LS contributed equally to this work.

\subsection{Data availability} The data that support the findings of this work are openly available \cite{Zenodo}.


\begin{thebibliography}{10}

\bibitem{Stanleybook}
Stanley, H. E. \emph{Introduction to Phase Transitions and Critical Phenomena}. (Oxford Science Publications, New York, 1971).

\bibitem{prb2025}
Soares, S. M., Squillante, L., Lima, H. S., Tsallis, C. \& de Souza, M. Universally nondiverging Gr\"uneisen parameter at critical points. \emph{Phys. Rev. B} \textbf{111}, L060409 (2025).

\bibitem{tsallisbook}
Tsallis, C. \emph{Introduction to Nonextensive Statistical Mechanics---Approaching a Complex World}. (Springer, New York, 2009). 2nd ed. (Springer, Switzerland, 2023).

\bibitem{ergodic}
Mac\'e, N., Alet, F. \& Laflorencie, N. Multifractal scalings across the many-body localization transition. \emph{Phys. Rev. Lett.} \textbf{123}, 180601 (2019).

\bibitem{2q}
Tsallis, C. Generalization of the possible algebraic basis of \emph{q}-triplets. \emph{Eur. Phys. J. Special Topics} \textbf{226}, 455 (2017).

\bibitem{vonNeumann}
von Neumann, J. \emph{Mathematical Foundations of Quantum Mechanics}. (Princeton University Press, New Jersey, 1955).

\bibitem{vonNeumannarticle}
von Neumann, J. Thermodynamik quantenmechanischer Gesamtheiten. \emph{G\"ott. Nachr. S.} \textbf{1927}, 273 (1927).

\bibitem{supercooled}
Gomes, G. O., Stanley, H. E. \& de Souza, M. Enhanced Gr\"uneisen parameter in supercooled water. \emph{Sci. Rep.} \textbf{9}, 12006 (2019).

\bibitem{griffiths}
Mello, I. F., Squillante, L., Gomes, G. O., Seridonio, A. C. \& de Souza, M. Griffiths-like phase close to the Mott transition. \emph{J. Appl. Phys.} \textbf{128}, 225102 (2020).

\bibitem{quantumGruneisen}
Squillante, L., Ricco, L. S., Ukpong, A. M., Lagos-Monaco, R. E., Seridonio, A. C. \& de Souza, M. Gr\"uneisen parameter as an entanglement compass and the breakdown of the Hellmann-Feynman theorem. \emph{Phys. Rev. B} \textbf{108}, L140403 (2023).

\bibitem{sabine}
Harbach, U. \& Hossenfelder, S. The Casimir effect in the presence of a minimal length. \emph{Phys. Lett. B} \textbf{632}, 379 (2006).

\bibitem{hess}
Fischbein, E., Tirosh, D. \& Hess, P. The intuition of infinity. \emph{Educ. Stud. Math.} \textbf{10}, 3 (1979).

\bibitem{infinity2025}
Mageed, I. A. The enduring enigma: philosophical and foundational challenges to infinity in mathematics and science. \emph{Optimality} \textbf{2}, 23 (2025).

\bibitem{osterloh}
Osterloh, A., Amico, L., Falci, G. \& Fazio, R. Scaling of entanglement close to a quantum phase transition. \emph{Nat.} \textbf{416}, 608 (2002).

\bibitem{kitaev}
Vidal, G., Latorre, J. I., Rico, E. \& Kitaev, A. Entanglement in Quantum Critical Phenomena. \emph{Phys. Rev. Lett.} \textbf{90}, 227902 (2003).

\bibitem{korepin}
Its, A. R., Jin, B.-Q. \& Korepin, V. E. Entanglement in the XY spin chain. \emph{J. Phys. A} \textbf{38}, 2975 (2005).

\bibitem{tsallistop}
Tsallis, C. Possible generalization of Boltzmann-Gibbs statistics. \emph{J. Stat. Phys.} \textbf{52}, 479 (1988).

\bibitem{pfeuty}
Pfeuty, P. The one-dimensional Ising model with a transverse field. \emph{Ann. Phys.} \textbf{57}, 79 (1970).

\bibitem{multifractal}
Voliotis, D. Multifractality in aperiodic quantum spin chains. \emph{J. Phys. A: Math. Theor.} \textbf{52}, 475001 (2019).

\bibitem{spindependence}
Goli, V. M. L. D. P., Sahoo, S., Ramasesha, S. \& Sen, D. Quantum phases of dimerized and frustrated Heisenberg spin chains with $s=1/2, 1$ and $3/2$: an entanglement entropy and fidelity study. \emph{J. Phys.: Condens. Matter} \textbf{25}, 125603 (2013).

\bibitem{zettili}
Zettili, N. \emph{Quantum Mechanics: Concepts and Applications}. (Wiley, 2009).

\bibitem{sachdev}
Sachdev, S. \emph{Quantum Phase Transitions}. (Cambridge University Press, Cambridge, 2001).

\bibitem{sachdevIsing}
Sachdev, S. \& Keimer, B. Quantum Criticality. \emph{Phys. Today} \textbf{64}, 29 (2011).

\bibitem{cohen3}
Cohen-Tannoudji, C., Diu, B. \& Lalo{\"e}, F. \emph{Quantum Mechanics, Volume 3: Fermions, Bosons, Photons, Correlations, and Entanglement}. (Wiley-VCH, Weinheim, 2019).

\bibitem{TsallisCaruso}
Caruso, F. \& Tsallis, C. Nonadditive entropy reconciles the area law in quantum systems with classical thermodynamics. \emph{Phys. Rev. E} \textbf{78}, 021102 (2008).

\bibitem{spin1}
Benyoussef, A. \& Ez-Zahraouy, H. Magnetic properties of a transverse spin-1 Ising model with random crystal-field interactions. \emph{J. Phys.: Condens. Matt.} \textbf{6}, 3411 (1994).

\bibitem{TsallisAndre}
Souza, A. M. C., Rap\v{c}an, P. \& Tsallis, C. Area-law-like systems with entangled states can preserve ergodicity. \emph{Eur. Phys. J. Special Topics} \textbf{229}, 759 (2020).

\bibitem{sakurai}
Sakurai, J. J. \& Napolitano, J. \emph{Modern Quantum Mechanics}, 2nd ed. (Addison-Wesley Publishing Company, San Francisco, 1994).

\bibitem{matrixrepresentation}
Catarina, G. \& Murta, B. Density-matrix renormalization group: a pedagogical introduction. \emph{Eur. Phys. J. B} \textbf{96}, 111 (2023).

\bibitem{lebellac}
Le Bellac, M. \emph{Quantum Physics}. (Cambridge University Press, Cambridge, 2006).

\bibitem{hilbertspace}
Cohen, D. W. \emph{An Introduction to Hilbert Space and Quantum Logic}. (Springer-Verlag, New York, 1989).

\bibitem{NNN}
Guimar\~aes, P. R. C., Plascak, J. A., Barreto, F. C. S. \& Florencio, J. Quantum phase transitions in the one-dimensional transverse Ising model with second-neighbor interactions. \emph{Phys. Rev. B} \textbf{66}, 064413 (2002).

\bibitem{schmidtoriginal}
Schmidt, E. Zur Theorie der linearen und nichtlinearen lntegralgleichungen. \emph{Math. Ann.} \textbf{63}, 433 (1907).

\bibitem{quantumcomputation}
Nielsen, M. A. \& Chuang, I. L. \emph{Quantum Computation and Quantum Information}, 10th ed. (Cambridge University Press, Cambridge, 2011).

\bibitem{entanglementschmidt}
Zhang, C., Denker, S., Asadian, A. \& G\"uhne, O. Analyzing Quantum Entanglement with the Schmidt Decomposition in Operator Space. \emph{Phys. Rev. Lett.} \textbf{133}, 040203 (2024).

\bibitem{thermoschmidt}
Malavazi, A. H. A. \& Brito, F. A Schmidt Decomposition Approach to Quantum Thermodynamics. \emph{Entropy} \textbf{24}, 1645 (2022).

\bibitem{schmidtblackhole}
Belokolos, E. D. \& Teslyk, M. V. Scalar field entanglement entropy of a Schwarzschild black hole from the Schmidt decomposition viewpoint. \emph{Class. Quantum Grav.} \textbf{26}, 235008 (2009).

\bibitem{chaostsallis}
Tsallis, C. Chaos. \emph{Chaos} \textbf{35}, 103131 (2025).

\bibitem{vidalprl}
Vidal, G. Efficient Classical Simulation of Slightly Entangled Quantum Computations. \emph{Phys. Rev. Lett.} \textbf{91}, 14 (2003).

\bibitem{quantumcoding}
Schumacher, B. Quantum Coding. \emph{Phys. Rev. A} \textbf{51}, 4 (1995).

\bibitem{quantumrefrigerator}
Levy, A., Alicki, R. \& Kosloff, R. \emph{Phys. Rev. E} \textbf{85}, 061126 (2012).

\bibitem{griffithsbook}
Griffiths, D. J. \& Schroeter, D. F. \emph{Introduction to Quantum Mechanics}, 3rd ed. (Cambridge University Press, Cambridge, 2018).

\bibitem{HSinfinity}
Gieres, F. Mathematical surprises and Dirac's formalism in quantum mechanics. \emph{Rep. Prog. Phys.} \textbf{63}, 1893 (2000).

\bibitem{fragmentation}
Adler, D., Wei, D., Will, M., Srakaew, K., Agrawal, S., Weckesser, P., Moessner, R., Pollmann, F., Bloch, I. \& Zeiher, J. Observation of Hilbert space fragmentation and fractonic excitations in 2D. \emph{Nat.} \textbf{636}, 80 (2024).

\bibitem{SbSI}
Tatsuzaki, I., Itoh, K., Ueda, S. \& Shindo, Y. Strain Along \emph{c} Axis of SbSI caused by Illumination in dc Electric Field. \emph{Phys. Rev. Lett.} \textbf{17}, 4 (1966).

\bibitem{SbSI1962}
Fatuzzo, E., Harbeke, G., Merz, W. J., Nitsche, R., Roetschi, H. \& Ruppel, W. Ferroelectricity in SbSI. \emph{Phys. Rev.} \textbf{127}, 2036 (1962).

\bibitem{electricdipole}
Canu, G., Bottaro, G., Buscaglia, M. T., Costa, C., Condurache, O., Curecheriu, L., Mitoseriu, L., Buscaglia, V. \& Armelao, L. Ferroelectric order driven Eu$^{3+}$ photoluminescence in BaZr$_x$Ti$_{1-x}$O$_3$. \emph{Sci. Rep.} \textbf{9}, 6441 (2019).

\bibitem{unveiling}
Squillante, L., Mello, I. F., Gomes, G. O., Seridonio, A. C., Lagos-Monaco, R. E., Stanley, H. E. \& de Souza, M. Unveiling the physics of the Mutual interactions in paramagnets. \emph{Sci. Rep.} \textbf{10}, 7981 (2020).

\bibitem{Zenodo}
Soares, S. M., Squillante, L., Lima, H. S., Tsallis, C. \& de Souza, M. Data for the paper entitled Universal and non-universal facets of quantum critical phenomena unveiled along the Schmidt decomposition theorem. Zenodo 10.5281/zenodo.17903468 (2025).


\end{thebibliography}
\end{document}